# Detectors for Energy-Resolved Fast Neutron Imaging

V. Dangendorf, C. Kersten and G. Laczko

*Physikalisch Technische Bundesanstalt, Braunschweig, Germany*

D. Vartsky, I. Mor, M.B. Goldberg and G. Feldman

*Soreq NRC, Yavne, Israel*

A. Breskin and R. Chechik

*Weizmann Institute of Science, Rehovot, Israel*

O. Jagutzki and U. Spillman

*University of Frankfurt, Frankfurt, Germany*

**\* Corresponding author:**

V. Dangendorf, Tel: +49 531 592 7525; fax: +49 531 592 7205; e-mail: volker.dangendorf@ptb.de

**Abstract**

Two detectors for energy-resolved fast-neutron imaging in pulsed broad-energy neutron beams are presented. The first one is a neutron-counting detector based on a solid neutron converter coupled to a gaseous electron multiplier (GEM). The second is an integrating imaging technique, based on a scintillator for neutron conversion and an optical imaging system with fast framing capability.

## 1. Introduction

The possibility of exploiting the characteristic resonant structure of the fast-neutron cross-section energy curves for different elements has led to the development of an element-specific, fast-neutron radiography and tomography method [1,2]. This method (Fast-Neutron Resonance Transmission-FNRT) utilizes the modification of a broad energy neutron spectrum transmitted through an inspected object due to resonant features in the cross-section of elements present. The object is illuminated by a pulsed neutron beam of broad-energy distribution (0.8 - 10 MeV) and the energy dependence of neutron transmission through an object is measured via the time-of-flight (TOF) technique. This method has been applied to the detection of elements such as C, O, N and H for determining the composition of agricultural products and for detecting contraband [3,4]. A system for detecting of explosives in passenger bags based on this method has also been constructed and tested [5-7].

The requirements for a neutron detector for FNRT applications are as follows: 1) It must be an imaging device (i.e. providing X-Y position coordinates of detected neutrons; 2) It must perform neutron spectroscopy with adequate energy resolution; 3) It must detect fast neutrons efficiently over a broad energy range; 4) It must have a large sensitive area for imaging bulky objects (> 20 x 20 cm$^2$) and 5) It must operate at high counting rates (>10$^6$ s$^{-1}$).

High-speed arrays of detectors for contraband identification using FNRT have been proposed by Van Staagen at al. [7] and Gibson et al. [8]. They consist of a matrix of individual scintillation detectors configured as a two-dimensional array. Another arrangement for an X-Y neutron detector for FNRT has been proposed by Miller [9]. The position resolutions obtained with the above methods were of the order of a few cm. They permitted reliable detection of bulk explosives, but were not sufficient for detection of thin objects. An alternative to the above detectors has been proposed by the MIT group [10] for operation using discrete-energy neutrons, mapping one element at a time. Here the detector was a scintillator viewed by a CCD camera using appropriate optics. The spatial resolution obtained was in the sub-millimetre range.

Recently we have proposed the possibility of performing FNRT in a pulsed, broad energy fast-neutron beam at the neutron facility of PTB [11]. Neutrons are produced by a nanosecond pulsed deuterium beam hitting a thick Be target. Two different imaging devices were developed: One is a pulse-counting detector based on a hydrogenous neutron converter coupled to a position-sensitive gaseous detector. The second is based on a fast plastic scintillator screen viewed by gated intensified CCD cameras. Both systems are capable of capturing images with about one millimetre spatial resolution and < 10 ns TOF resolution.

## 2. Neutron Counting Gaseous Detector

The FAst Neutron GASeous (FANGAS) imaging detector is based a on a hydrogen-rich radiator (e.g., polypropylene (PP)) for fast-neutron to proton conversion. Some of the protons escape from the radiator and are detected by the adjacent charged-particle detector. Fig. 1 shows a schematic drawing of the device. The thickness of the radiator is 1 mm, which is a compromise between the requirement of a thick converter for high neutron absorption and the limited escape length for protons from PP.

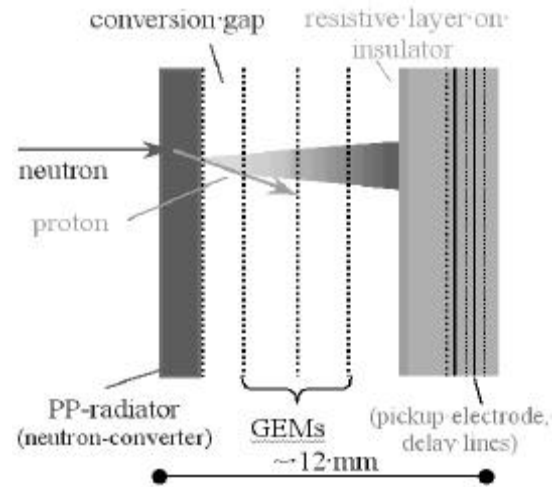

Fig. 1: Schematic drawing of the fast-neutron gaseous detector FANGAS

The calculation and optimisation of the radiator efficiency was performed with GEANT [12]. For a 1 mm thick converter the calculated efficiency for 5 MeV neutrons is 0.2 %. The recoil protons ionise the gas in the thin gap between the foil and the electron amplifier. The ionisation electrons are collected and multiplied in an electron multiplier consisting of three cascaded GEMs. Leaving the last GEM and passing the induction gap, the final charge avalanche is collected on a resistive Ge layer. The propagating charge in the induction gap induces a well-localised signal on a position encoding pickup electrode. This electrode is located about 3 mm behind the resistive anode, which causes the induced signal on the pickup electrode to be spread over several millimetres. The position read out is effected by this double-sided pad-structured pickup electrode and the delay-line technique. This method was already applied earlier by one of us to the readout of position-sensitive photon detectors [13]. A recent study of the resistive anode technique can be found in [14].

In the present prototype detector we are using GEMs made at CERN of 100 x 100 mm size. The gas filling is Ar(70%)/CO$_2$ (30%). With this detector we have made radiographic measurements at the PTB neutron facility, using a 13.3 MeV deuteron beam on a 3 mm thick Be target, yielding a broad-energy neutron beam.

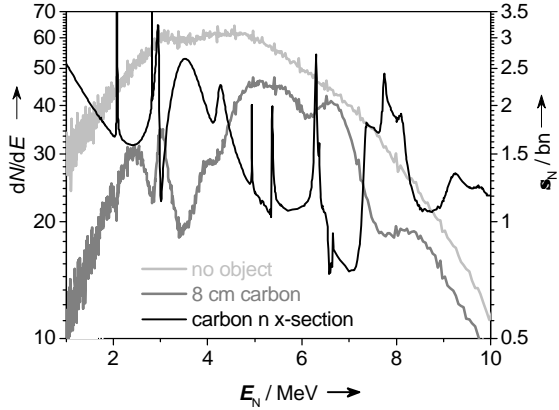

Fig. 2: Energy spectra, derived from TOF measurements with the FANGAS detector. The grey curves are without and with a 8 cm thick carbon sample. The black curve shows the total neutron cross section of carbon. The pronounced broad minimum (at 6,5 – 7,2 MeV) and maximum (at 7,3 – 8,2 MeV) which are used for demonstrating the method of differential radiography are clearly reflected in the measured spectra.

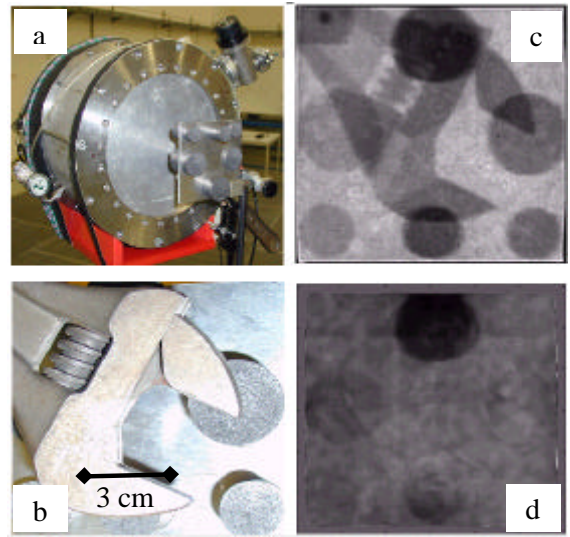

Fig. 3: Imaging of carbon rods and a steel wrench with fast neutrons: . a) detector and sample in the neutron beam, b) the sample which consists of a steel wrench and carbon rods of 30 and 20 mm in diameter and 20, 40 and 60 mm length, c) and d) show neutron images (see text).

Details of the facility and the beam properties, as well as test results with an earlier neutron-sensitive wire chamber, are reported elsewhere [11]. Fig. 2 shows the energy spectra, derived from the TOF spectra measured without a sample and with an 8 cm thick C-block. For comparison, the total neutron cross-section for carbon is shown. Fig 3a shows the neutron detector with a sample made of carbon rods of various sizes and a steel wrench. While carbon has a pronounced minimum and maximum in the neutron cross-section between 6.5 and 8.5 MeV, the cross-section for steel is flat. By taking an image ON and OFF the resonance energy and dividing these images pixel by pixel, the steel wrench should thus disappear and only the carbon remains visible. .

Figs. 3c and 3d show the results obtained by irradiating the sample shown in 3b. The neutron transmission image taken with the broad spectrum between 1 and 10 MeV shows the wrench and the carbon rods. The neutron images are corrected with a flat-field reference spectrum obtained without sample. The ON/OFF resonance ratio is shown in Fig. 3d. As expected, the wrench disappears. However, due to the weak contrast of the thinner carbon samples and the poor pixel statistics, the smaller carbon samples are also not visible in the ratio image.

## 3. Scintillator with Intensified CCD Camera

The second detector, the OpTIcal Fast-NeuTron Imaging system (OTIFANTI), is schematically shown in Fig. 4. Its front-end consists of a fast plastic scintillator (BC400), 220x220x10mm$^3$ in dimensions. Neutrons interact in the plastic and transfer part of their energy to a proton. The scintillation light is projected via a front-side coated Al mirror and a large-aperture lens to an optical detection system. The optical system that registers the scintillations is required to collect light produced by neutrons of a selected energy window (a few hundred keV broad) out of the overall energy distribution. In a pulsed neutron beam this corresponds to a well-defined time window of a few ns width at a specific delay relative to the start time of the neutron burst. The pulse width of the deuteron beam as well as the decay constant of the scintillation light need to be small compared to the TOF difference corresponding to the two required neutron energies. This is the main reason for the choice of a fast plastic scintillator despite its relatively poor scintillation efficieny compared to commercial ZnS-based scintillators for fast neutrons

The optical detection system requires a very fast exposure control of the CCD camera. The exposure time window per neutron pulse is typically of the order of 5 – 20 ns, the pulse repetition rate is 1–2 MHz and the total exposure time for an image is of the order of tens of seconds. This is accomplished by means of a gated image intensifier used as a fast electronic shutter in front of the CCD. The present optical system suffers from a rather low optical detection efficiency of the scintillation light. This has the effect that, on the average, only one out of five neutrons (at 5 MeV), which interact in the scintillator are detected. This leads to an overall neutron detection efficiency of 1.25 %. Experimental results obtained with this instrument will be published in a separate article.

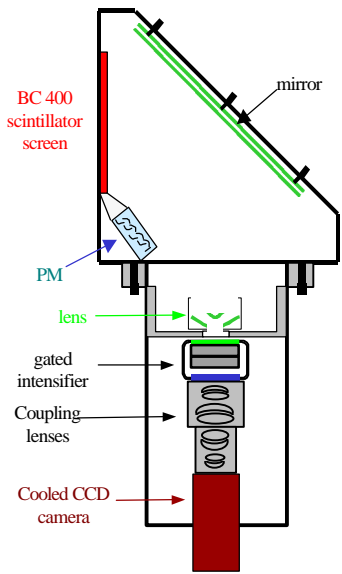

Fig. 4: Schematic drawing of the optical fast-neutron detector OTIFANTI

## 4. Outlook

In this chapter we will briefly describe the future development, which we consider as necessary for applying these detectors in industrial FNRT.

The present gaseous detector has a rather small size which is not suitable for practical use in FNRT. Detectors using 30x30 $cm^2$ GEMs have been produced and are successfully being operated [15]. To overcome the very low neutron detection efficiency, we plan to cascade 25 detector elements along the beam axis, which would provide a reasonable detection efficiency of 5 %.

The optical detector, as mentioned above, suffers from low light collection efficiency of the scintillation light. The following improvements are in course:

- a lens with a lower f-number (f/1.0) will provide a factor of 3.3 higher optical efficiency than the present one.
- a larger intensifier (∅ 40 mm) will increase the optical efficiency by another factor of 2.5
- a thicker scintillating fibre slab will increase the neutron detection efficiency and improve position resolution

V. D. acknowledges the support of the Minerva foundation. A.B. is the W. P. Reuther Professor of Research in the Peaceful use of Atomic Energy.